\documentclass[%
 reprint,
 amsmath,amssymb,
 aps,
]{revtex4-2}

\usepackage{graphicx}
\usepackage{dcolumn}
\usepackage{bm}
\usepackage{multirow}

\newcommand{\aj}{Astron. J.}

\begin{document}

\preprint{APS/123-QED}

\title{Learning Intrinsic Alignments from Local Galaxy Environments}

\author{Matthew Craigie}
\affiliation{School of Mathematics and Physics, University of Queensland, QLD 4072, Australia}

\author{Eric Huff}%
\affiliation{Jet Propulsion Laboratory, California Institute of Technology, Pasadena, CA 91109, USA}

\author{Yuan-Sen Ting} 
\affiliation{Department of Astronomy, The Ohio State University, Columbus, OH 43210, USA}
\affiliation{Center for Cosmology and AstroParticle Physics (CCAPP), The Ohio State University, Columbus, OH
43210, USA}

\author{Rossana Ruggeri}
\affiliation{School of Physics and Chemistry,  Queensland University of Technology, QLD 4000, Australia}
\affiliation{School of Mathematics and Physics, The University of Queensland, QLD 4072, Australia}

\author{Tamara M.\ Davis}
\affiliation{School of Mathematics and Physics, The University of Queensland, QLD 4072, Australia}

\date{\today}

\begin{abstract}
We present DELTA (Data-Empiric Learned Tidal Alignments), a deep learning model that isolates galaxy intrinsic alignments (IAs) from weak lensing distortions using only observational data. The model uses an Equivariant Graph Neural Network backbone suitable for capturing information from the local galaxy environment, in conjunction with a probabilistic orientation output. Unlike parametric models, DELTA flexibly learns the relationship between galaxy shapes and their local environments, without assuming an explicit IA form or relying on simulations. When applied to mock catalogs with realistic noisy IAs injected, it accurately reconstructs the noise-free, pure IA signal. Mapping these alignments provides a direct visualization of IA patterns in the mock catalogs. Combining DELTA with deep learning interpretation techniques provides further insights into the physics driving tidal relationships between galaxies. This new approach to understanding and controlling IAs is suitable for application to joint photometric and spectroscopic surveys such as the combination of upcoming Euclid, Rubin, and DESI datasets.
\end{abstract}

\maketitle

\section{Introduction}\label{sec:introduction}
The structure of matter on cosmic scales encodes a wealth of information about the universe's properties and evolution. One of the most powerful modern probes of this large-scale structure is weak lensing, the subtle distortion of galaxy shapes due to the gravitational influence of intervening matter. Weak lensing is highlighted as a primary science objective for various upcoming missions including the Euclid Space Telescope \citep{euclid2024}, Nancy Grace Roman Space Telescope \citep{roman2015}, and Vera C. Rubin Observatory \citep{rubin2019}. Together, these experiments promise to deliver unprecedented constraints on cosmological models.

One of the primary challenges in upcoming weak lensing studies is contamination from galaxy intrinsic alignments (IAs), which is the tendency of nearby galaxies to align their orientations due to tidal interactions \citep{croft2000, heavens2000, lee2000, lee2001, catelan2001}. These alignments can mimic the weak lensing signal, introducing biases and degrading the precision of cosmological constraints derived from lensing analyses \citep{joachimi2010, joachimi2011, kirk2012, krause2015,  blazek2019, desy3shear, campos2023}. 

As a result, the lensing community has devoted substantial effort to modeling and mitigating IAs. However, modeling IAs is difficult due to their strong observational coupling with gravitational lensing. Existing methods for disentangling these effects leverage the distinct origins of these signals: IAs arise from local tidal fields, whereas weak lensing is a result of distant gravitational influence along the line of sight. Techniques exploit this difference by selecting galaxy pairs that are close in redshift \citep{hirata2007, singh2015, johnston2019}, subtracting a modeled lensing signal based on foreground structure \citep{mandelbaum2013}, or applying nulling methods that exploit the differing radial dependencies of the two signals \citep{blazek2012}. However, all of these approaches are limited to measuring two-point correlations of galaxy shapes. As a result, they cannot capture the full complexity of IA patterns and structures, which are a result of non-linear physics that requires higher-order statistics to fully characterize.

Due to the observational difficulty of isolating IAs, current models rely on physical assumptions about galaxy formation and interactions to construct parameterized analytic forms for IAs. These models typically include a small number of free parameters that are marginalized over in a full weak lensing analyses. The predominant IA model in contemporary lensing studies \citep[e.g.][]{desy3shear} is the Tidal Alignment and Tidal Torquing (TATT) model \citep{blazek2019}. TATT consists of two physically motivated components: a tidal alignment mechanism, which causes the diffuse halos of elliptical galaxies to align with nearby mass distributions, and a tidal torquing mechanism, which imparts angular momentum to spiral galaxies, preferentially aligning their discs.

Despite its widespread use, the TATT model faces several challenges. While tidal alignments in elliptical galaxies are well supported by observations \citep{mandelbaum2006, hirata2007, joachimi2011, blazek2011, singh2015, tsaprazi2022}, observational studies of tidal torquing in spiral galaxies report mixed results: some detect evidence for the effect \citep{zhang2015}, while others find no significant signal \citep{pahwa2016, johnston2019, desy3shear, desy3cross}. Similarly, studies using hydrodynamical simulations produce conflicting results: some detect no evidence of alignment \citep{samuroff2021}, others report strong signatures \citep{delgado2023}, and some observe alignments only at high redshifts \citep{zjupa2020}. These inconsistencies across both observations and simulations point to a potential mismatch between current modeling assumptions and the underlying physics of alignments.


A key contributor to this mismatch is the limited expressiveness of existing models, which often fail to capture the complexity of the astrophysical processes driving intrinsic alignments. For example, simple parametric forms such as power-law redshift evolution do not reproduce the observed IA evolution, even for elliptical galaxies \citep{georgiou2019, desy3shear, chen2024}. At small scales, where nonlinear processes become significant, current models also struggle to predict alignment behavior accurately \citep{georgiou2019, johnston2019}. As next-generation weak lensing surveys extend to higher redshifts and smaller physical scales \citep[e.g.][]{aric2023, gordon2024}, IA modeling must evolve accordingly to fully exploit the increased statistical power \citep{georgiou2019, samuroff2021, desy3shear}.

Addressing the challenges of intrinsic alignment modeling is essential to realize the full potential of the Euclid, Roman, and Rubin weak lensing surveys. These experiments will produce datasets with unprecedented statistical precision, leaving systematic uncertainties such as IAs to dominate \citep{krause2015}. Without improved IA models, analyses risk introducing biases or underestimating uncertainties, thereby limiting the accuracy and reliability of future cosmology measurements.

To overcome current modeling limitations, we propose a data-driven deep learning framework: Data-Empiric Learned Tidal Alignments (DELTA). Like previous methods, DELTA separates intrinsic alignments from lensing distortions by exploiting their distinct physical origins. However, instead of assuming an explicit form of IAs, it uses a deep learning model to implicitly infer the relationship between a galaxy’s shape and the surrounding galaxy distribution. By restricting model inputs to spatially local information, DELTA captures the tidally-induced IAs while remaining insensitive to lensing distortions, which depend on more distant effects. With this approach, DELTA learns directly from observational data and therefore does not require any simulations or forward modeling.

DELTA serves two primary purposes. First, it can replace traditional alignment models in weak lensing analyses to mitigate IA contamination and improve the precision of cosmological constraints. Second, DELTA provides a method for better understanding the underlying physical relationships between galaxy shape and the local matter distribution. This enables the development of more comprehensive IA models and provides insights into the complex, and still poorly understood, processes governing galaxy evolution.

In this work, we demonstrate DELTA's capabilities by applying it to a simulated galaxy distribution with a realistic injected intrinsic alignment signal. We show that DELTA successfully isolates the IA signal, even in the presence of realistic shape noise. The model recovers a non-trivial, nonlinear alignment structure without relying on any physical assumptions beyond global isotropy and homogeneity. Additionally, we apply two deep learning interpretation techniques to characterize the learned alignments and recover physical insights consistent with the injected signal. These results highlight DELTA's potential as a powerful tool for modeling and interpreting intrinsic alignments in observational data, with direct implications for improving cosmological constraints from upcoming surveys.

\section{Recovering Intrinsic Alignments with DELTA}\label{sec:delta}
The Data-Empiric Learned Tidal Alignments (DELTA) model separates intrinsic alignments from lensing by considering their distinct physical origins: short-range tidal forces and long-range gravitational lensing, respectively. DELTA infers the tidal field directly from cosmological survey data, providing predictions of the tidally induced shape for each galaxy. In this section, we present the theoretical foundations of DELTA, outline its implementation, and demonstrate its application to a realistic simulated dataset.

\subsection{Theoretical Foundations}\label{sec:theoretical_foundations}

In weak lensing studies, a galaxy's shape is commonly represented by a complex ellipticity $\epsilon$, where the real and imaginary components correspond to the degree of elongation in the $+$ and $\times$ directions, respectively. In the weak lensing regime, the ellipticity can be decomposed as the sum of three components: 
\begin{equation}
    \epsilon = \epsilon_T + \epsilon_L + \epsilon_N.
\end{equation}
Here, $\epsilon_T$ is the tidally induced component, representing shape distortions due to local tidal interactions. $\epsilon_L$ is the lensing-induced component, arising from distortions caused by foreground gravitational lensing. $\epsilon_N$ is the shape noise, reflecting stochastic variations in galaxy shapes due to the random initial conditions of the universe. This noise dominates the signal, typically contributing more than $95\%$ of the total ellipticity magnitude. Of the remaining $<5\%$, the lensing component $\epsilon_L$ accounts for most of the signal, with the tidally induced component $\epsilon_T$ contributing roughly $1\%$. While these proportions vary with galaxy type, environment, and redshift, in all cases the observed ellipticity is dominated by noise.

The goal of DELTA is to isolate the tidal component of shape, $\epsilon_T$, from the lensing and noise components. We do this by considering how each component relates to cosmological survey observables:
\begin{itemize}
    \item The tidal component $\epsilon_T$ arises from the stretching of visible matter in the galaxy due to tidal forces experienced over its lifetime. These tidal forces are generated by local gravitational fields, which are themselves induced by the surrounding matter density. As a result, $\epsilon_T$ is intrinsically linked to the local galaxy distribution through their shared dependence on the local matter field. Because tidal forces scale with the inverse cube of distance, they are predominantly local in nature, with simulations showing that tidal alignments vanish beyond $\sim10$ Mpc \citep{zjupa2022, codis2015}.

    \item The lensing component $\epsilon_L$ arises from the deflection of light due to gravitational distortions of spacetime as light from the source galaxy travels to the observer. These distortions are induced by the matter distribution along the entire line of sight. Like tidal forces, lensing depends on the matter field, but in this case it is primarily related to foreground galaxies. Consequently, $\epsilon_L$ has negligible dependence on the local galaxy distribution.

    \item The noise component $\epsilon_N$ reflects stochastic variations in galaxy shapes caused by random initial conditions in the early universe, as well as by other unobservable physical processes. This component lacks any predictable structure and is uncorrelated with any cosmological survey observable.
\end{itemize}

Using these physical connections, we construct a deep learning model that predicts a galaxy's shape. To make predictions, the model is provided only with local galaxy positions and properties, which we denote as the galaxy's neighborhood $\mathcal{N}$. Since $\epsilon_T$ and $\mathcal{N}$ share a relationship, the model can use information from the neighborhood to predict $\epsilon_T$. In contrast, $\epsilon_L$ and $\epsilon_N$ have no relationship with $\mathcal{N}$, so the model has no information from which to predict them. Because both components are globally isotropic, they contribute only a random component to $\epsilon$ for any individual galaxy, irrespective of $\mathcal{N}$. Therefore, a model trained to predict the total ellipticity $\epsilon$ using only $\mathcal{N}$ can only predict the $\epsilon_T$ component. 


More formally, consider the conditional probability distribution of ellipticity given the neighborhood, $P(\epsilon | \mathcal{N})$. Assuming that the components of ellipticity are independent, the expected value of this distribution is
\begin{equation}
    \mathbb{E}[\epsilon | \mathcal{N}] = \mathbb{E}[\epsilon_T | \mathcal{N}] + \mathbb{E}[\epsilon_L | \mathcal{N}] + \mathbb{E}[\epsilon_N | \mathcal{N}].
\end{equation}

We assume that the neighborhood $\mathcal{N}$ contains no information correlated with lensing, such that $P(\epsilon_L | \mathcal{N}) = P(\epsilon_L)$. We discuss the validity and implications of this assumption in Section~\ref{sec:delta_discussion}. By definition, the noise component is also independent of the neighborhood: $P(\epsilon_N | \mathcal{N}) = P(\epsilon_N)$. Under the assumption of a statistically isotropic universe, lensing has no preferred direction, and therefore
\begin{equation}
    \mathbb{E}[\epsilon_L | \mathcal{N}] = \mathbb{E}[\epsilon_L] = 0,
\end{equation}
and similarly, the noise has zero mean:
\begin{equation}
    \mathbb{E}[\epsilon_N | \mathcal{N}] = \mathbb{E}[\epsilon_N] = 0.
\end{equation}

It follows that the expected value of the ellipticity conditioned on the neighborhood is equal to the expected value of the tidal component alone:
\begin{equation}\label{eq:expected_tidal}
    \mathbb{E}[\epsilon | \mathcal{N}] = \mathbb{E}[\epsilon_T | \mathcal{N}].
\end{equation}

When measuring alignments, it is useful to isolate the direction of elongation, which carries alignment information, from the magnitude of elongation, which does not. We express the ellipticity in polar form,
\begin{equation}
    \epsilon = |\epsilon|e^{2i\phi},
\end{equation}
where $|\epsilon|$ denotes the magnitude of ellipticity, corresponding to the degree of elongation, and $\phi\in[0, 2\pi]$ is the orientation angle, indicating the direction of elongation. The factor of 2 in the exponent captures the spin-2 rotational symmetry of ellipses.

Following from Equation~\ref{eq:expected_tidal}, we consider the expected value of the orientation angle:
\begin{equation}
    \mathbb{E}[\phi | \mathcal{N}] = \mathbb{E}[\phi_T | \mathcal{N}],
\end{equation}
where $\phi_T$ denotes the orientation angle associated with the tidal alignment component. This relation follows from the same reasoning as for $\epsilon$: both the lensing-induced angle $\phi_L$ and the noise component $\phi_N$ are globally isotropic, so their contributions vanish with an expected value. Consequently, $\phi_L$ is absorbed into the noise term. Without access to line-of-sight information, it contributes only a random orientation for each galaxy.

This theoretical framework enables a completely data-driven approach to intrinsic alignment modeling. Unlike traditional methods that require explicit parametric forms or assumptions about the physical mechanisms driving alignments, DELTA can learn the relationship $\text{E}[\phi_T|\mathcal{N}]$ directly from observational data without any prior model of tidal forces. The conditioning framework ensures that learned patterns must arise from correlations between local galaxy environments and intrinsic orientations, rather than from lensing or noise. This represents a completely distinct approach from previous methods, shifting from theoretical physical modeling to empirical pattern recognition. 


\subsection{The DELTA Model}
\begin{figure*}
    \centering
    \includegraphics[width=1.0\linewidth]{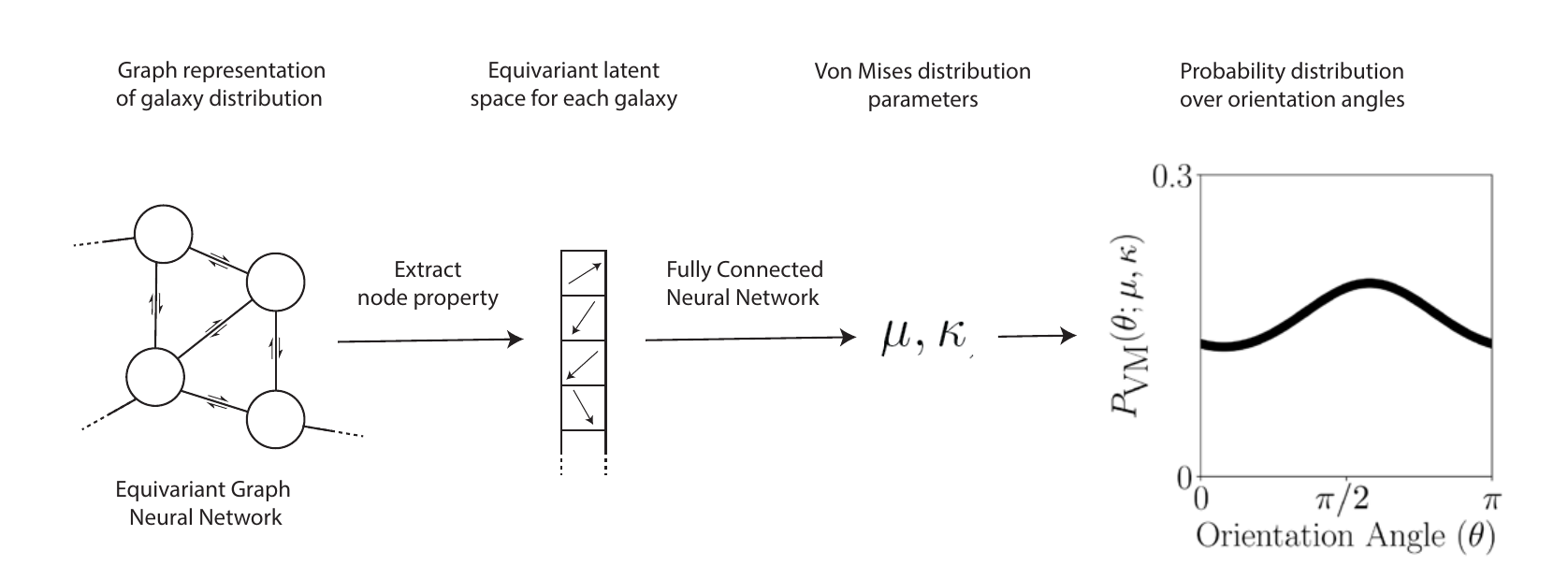}
    \caption{An overview schematic of the Data-Empiric Learned Tidal Alignments (DELTA) model. DELTA uses an equivariant graph neural network (EGNN) to capture structural information from the large-scale structure, accumulating a latent representation on each galaxy node. This equivariant latent space is a set of vectors containing a compressed representation of the relative structure in each galaxy's neighborhood. A fully connected neural network then converts these representations into $\mu$ (mean) and $\kappa$ (precision) parameters of a von Mises distribution over 2D angles. The DELTA model's resulting output is a likelihood of the galaxy's observed orientation angle, learned exclusively from the local structure, where the mean angle is free from lensing or noise information.}
    \label{fig:schematic}
\end{figure*}

The DELTA model leverages the conditioning framework described in Section~\ref{sec:theoretical_foundations} to isolate the tidal component and recover intrinsic alignments. It relies solely on observational data, making no assumptions about the physical mechanisms underlying alignments, and instead infers the relevant physics directly from patterns in the data. At a high level, DELTA operates in two stages: compression and probabilistic prediction. The compression stage extracts relevant information from the galaxy distribution and condenses it into a feature vector for each galaxy. The prediction stage then uses this feature vector to estimate the parameters of a probability distribution over the galaxy's orientation. The resulting output distribution encodes both the preferred alignment direction and the expected variability for each galaxy. Figure~\ref{fig:schematic} shows a schematic overview of the model architecture.




\subsubsection{Model Implementation \& Architecture}

Intrinsic alignments contribute only a small fraction of each galaxy's shape (and therefore orientation), and are strongly subdominant to the stochastic noise component. As a result, DELTA operates as a probabilistic regression task with noisy labels, where the noise dominates the signal by a factor of approximately 100. This presents a significant challenge for deep learning, which we address through careful architectural design that introduces inductive biases -- prior assumptions about the data embedded into the model -- to improve convergence and predictive performance. 

We structure DELTA in two distinct stages, each incorporating specific inductive biases. The first stage, compression, efficiently extracts information from the galaxy distribution. This stage enforces global invariance to rotation and translation, reflecting the assumptions of isotropy and homogeneity in the universe. By embedding these symmetries into the model, we eliminate the need to learn them from data, enabling a more efficient internal representation and reducing the amount of training data required. The second stage performs probabilistic orientation prediction. It transforms the deterministic output of the compression stage into a probability distribution over orientations. This probabilistic representation smoothly handles the noisy nature of intrinsic alignment predictions. Furthermore, by adopting a firmly parameterized form for the distribution, we simplify the prediction task promoting improved convergence and restricting over-fitting relative to more freely parameterized probabilistic models.

We implement DELTA in PyTorch, and make the code publicly available on GitHub \cite{craigie_delta}.
\\

\noindent
\textbf{Equivariant Graph Neural Network Compression}

\noindent
The first stage in DELTA is a deep compression model that transforms the high-dimensional, noisy information surrounding each source galaxy into a low-dimensional but maximally informative latent representation relevant to tidal alignments.

To perform this compression, DELTA uses a Graph Neural Network (GNN), a deep learning architecture tailored for graph-structured data, where nodes carry features and edges encode relationships. GNNs are well suited to sparse, coordinate-based data and enable efficient, generalized aggregation of information across graph structures. These properties make them particularly effective for a range of large-scale structure applications, with demonstrated success in characterizing halos in cosmology simulations \citep{larson2024}, and simulation-based inference with the large-scale structure \citep{desanti2023, lehman2024, roncoli2024, desanti2025}.


We use a subclass of GNNs known as Equivariant Graph Neural Networks (EGNNs) \citep{satorras2022}. These models produce output vectors that are equivariant under spatial rotations: rotating the input by an angle $\theta_0$ results in an output rotated by the same angle. This property allows the EGNN to aggregate information in a way that is consistent with the global isotropy of galaxy orientations.

We adapt the EGNN implementation from \citep{satorras2022}, where the model predicts output vectors based on weighted sums of the relative position vectors between connected nodes. This construction ensures that rotations of the input produce corresponding transformations of the output. For our application, we modify the EGNN in several ways to accommodate the structure of our data:
\begin{itemize}
    \item We output the predicted equivariant vector directly, rather than using it to update node positions as in the original EGNN, which was designed to model particle dynamics. Since galaxies are static, positional updates are unnecessary.
    \item We modify the equivariance to a spin-2 equivariance in the $xy$-plane, reflecting the spin-2 symmetry of observed galaxy orientations. This ensures that the output vector remains invariant under a rotation of the neighborhood by $\pi$ in the $xy$-plane.
    \item We project the 3D output vector into 2D to match the 2D orientation information available from lensing surveys. The EGNN computations operate on the full 3D galaxy distribution, and we apply the projection only in the final step.
\end{itemize}
We detail the EGNN's architecture in Appendix~\ref{sec:appendix_a}. 

Within the EGNN, the $i$th galaxy is connected to its neighborhood $\mathcal{N}_i$ through three message-passing hops, where hops occur with each galaxy's $k=10$ nearest neighbors. This approach imposes a simple distance limit that is approximately uniform across the dataset and sufficient for this demonstration. In future, we can use a more explicit constraint with a dependence on distance, effectively creating a local `bubble' representing the galaxy's neighborhood. As an output, the EGNN processes this neighborhood information in a latent set of 2D vectors that are equivariant to rotations in the $xy$ plane. 
\\

\noindent
\textbf{Probabilistic Orientation Prediction}

\noindent
The second stage of the DELTA model is a probabilistic prediction model, which takes the deterministic latent space and transforms it into a probabilistic prediction for each galaxy's orientation. The choice of a probabilistic output serves three purposes: to enable smoother handling of noise, to create a more accurate representation of the inherently noisy galaxy orientations, and to allow a notion of uncertainty propagate to downstream inference tasks.

DELTA learns to approximate the conditional probability distribution over galaxy orientation, denoted $q(\phi | \mathcal{N}) \approx P(\phi | \mathcal{N})$, using the local neighborhood $\mathcal{N}$ as input. For each galaxy $i$, DELTA predicts an orientation $\phi_i$ drawn from the learned distribution $\phi_i \sim q(\phi | \mathcal{N}_i)$. In a successfully trained model, the expected value of the learned distribution matches the true conditional expectation:
\begin{equation}
    \mathbb{E}_{q}[\phi | \mathcal{N}] = \mathbb{E}[\phi | \mathcal{N}].
\end{equation}
Given that the compression stage already converts $\mathcal{N}_i$ into a set of latent vectors $\mathbf{z}_i$, the probability modeling stage in isolation must capture the conditional distribution $\phi_i\sim q(\phi |\mathbf{z}_i)$.

To model the resulting probability distribution, we approximate that the resulting total orientation of each galaxy will be drawn from a von Mises distribution, which closely resembles a Gaussian distribution wrapped onto the circle. We choose the von Mises distribution because it parameterizes both direction and spread for directional probability. For an orientation angle $\theta$, the von Mises distribution has a probability density function of
\begin{equation}\label{eq:von_mises}
    P_\text{VM}(\theta; \mu, \kappa) = \frac{e^{\kappa \cos(\theta - \mu)}}{2\pi I_0(\kappa)}
\end{equation}
where $I_0(\kappa)$ is the modified Bessel function of the first kind of order 0. The location parameter $\mu$ controls the direction of the peak, while the concentration parameter $\kappa$ controls the strength of the directional peak. To compensate for the spin-2 nature of orientation, we map the distribution to $\theta=2\phi$, such that $\theta\in[0, 2\pi]$ while $\phi\in[0, \pi]$. Predictions of galaxy orientation will typically have a low concentration parameter $\kappa$ (and therefore wide spread) due to the uncertainty in prediction from noise dominance, with the case of no directional bias $\kappa=0$ returning to a uniform distribution. 

DELTA processes the EGNN's learned equivariant feature vectors into scalar spin-2 directions in the $xy$-plane, 
\begin{equation}
    \alpha_{i} = \arctan2(\textbf{z}_{i,2}, \textbf{z}_{z,1}),
\end{equation}
before using these angle scalars to predict the von Mises distribution parameters for each node,
\begin{equation}
    \mu_i = f_\mu(\alpha_i)
\end{equation}
and 
\begin{equation}
    \kappa_i = f_\mu(\kappa_i)
\end{equation}
where $f_\mu$ and $f_\kappa$ are fully connected neural networks. 

Although these parameters primarily define a probability distribution, they also have a direct physical interpretation. The output mean $\mu_i$ represents the expected value of $q(\phi|\mathcal{N}_i)$. This is the orientation we would expect, on average, if thousands of galaxies in the dataset shared the same neighborhood configuration. However, DELTA’s deep learning backbone enables it to identify patterns such that it can accumulate information and make accurate predictions without requiring repeated instances of identical neighborhoods. The value of $\mu_i$ therefore reflects an aggregation of the system’s overall behavior, projected onto an individual galaxy based on its local environment.

The output concentration $\kappa_i$ characterizes the concentration (i.e., inverse spread) of the distribution $q(\phi|\mathcal{N}_i)$ and reflects the range of plausible alignments for each galaxy. It does not represent a direct uncertainty on $\mu_i$, but it is related to the model's confidence in its prediction. A high $\kappa_i$ indicates that the model has identified a strong directional signal from the neighborhood and is confident in the predicted alignment. In contrast, a low $\kappa_i$ suggests that the model is uncertain, either because it has not learned a reliable relationship between $\mathcal{N}$ and $\phi$, the neighborhood lacks sufficient information, or the galaxy’s orientation distribution is not well-modeled by a von Mises distribution. Both higher signal-to-noise in the alignments and larger training datasets tend to reduce $\kappa_i$, indicating more confident predictions.

This approach is similar to a Mixture Density Network \citep{bishop1994}, but uses a single distribution rather than a mixture. Limiting the output to a single, well-parameterized distribution reduces the number of degrees of freedom, which simplifies the prediction task and improves model interpretability. In contrast, a multi-peaked distribution would introduce additional complexity when interpreting the expected orientation. Additionally, by directly predicting $\phi$ in a von Mises distribution rather than the two components of $\epsilon$, we remove a degree of freedom in the shape which removes degeneracy and improves learning stability.

\subsection{Recovery of IAs in Mocks with DELTA}

We apply DELTA to a mock dataset designed to replicate the characteristics of observational data and demonstrate its ability to recover intrinsic alignments from noisy galaxy orientations.

\subsubsection{Mock Datasets}

We use a set of mock datasets that reproduce intrinsic alignments with properties similar in scale and structure to those in observational surveys. We construct the mocks following the method described in \cite{vanalfen2024}, using the \texttt{halotools} Python package. Briefly, this approach uses the MultiDark dark matter simulation \citep{prada2012}, populates halos with galaxies through a halo occupation distribution model \citep{behroozi2013}. These mocks provide 3D spatial positions for 2.27 million central and satellite galaxies within a 1~Gpc$^3$ cubic volume. 



To construct realistic alignment mocks, the method injects a `pure' intrinsic alignment signal as a 3D orientation vector and adds directional noise to simulate observational conditions. For central galaxies, the pure signal aligns each galaxy with the major axis of its host dark matter halo's ellipsoidal shape. For satellite galaxies, the signal enforces a radial alignment toward the halo center. The added noise draws from a Dimroth-Watson distribution, a directional probability distribution on the sphere that symmetrically distributes orientations about a preferred axis. Because the pure alignments match the alignment N-body dark matter halos, which reflect the fully non-Gaussian large-scale structure field to all orders, they encode higher-order information that cannot be captured solely by two-point statistics. 

We test three realistic alignment strength scenarios: low, medium, high. These correspond to the alignments measured in low, intermediate and high mass galaxies, as determined from hydrodynamical simulations \citep{vanalfen2024}. We do not assign mass-dependent alignments, but rather test as if all galaxies in the sample aligned with the corresponding strength of that mass group. In each scenario, satellite galaxy alignments are weaker than central galaxy alignments. In Table~\ref{tab:results} we provide each alignments scenario's corresponding alignment strength parameters. The specifics of the alignment parameter is described in \cite{vanalfen2024}, but in general, a lower alignment parameter indicates more shape noise and less clear alignment signal.  

We project these orientations into 2D unit vectors in the $xy$-plane to match the 2D shape information available in observational data, and convert them to spin-2 orientation angles $\phi$. For each galaxy, we also compute the corresponding spin-2 orientation for a full alignment scenario, which we denote by $\hat\phi$. These noise-free orientations represent the pure intrinsic alignment signal and are used during verification.

We do not explicitly inject a lensing signal. Since the model receives no line-of-sight conditioning, and each galaxy only has information about its own orientation during training, the lensing component for each galaxy is randomly oriented and indistinguishable from the existing isotropic shape noise. 

\subsubsection{Model Setup and Training}

The training dataset consists of galaxies indexed by $i$, each described by node positions and properties $(\vec{r}_i, \vec{p}_i)$, along with a target orientation $\phi_i$. In our main analysis, we do not include any galaxy observables, and instead set $\vec{p}_i = (1,)$ for all galaxies. This ensures that the model extracts alignment information solely from the galaxies' relative positions. For the interpretability analysis, we modify $\vec{p}_i$ to include nontrivial properties, as described in Section~\ref{sec:permutation_test}.

We divide the data along the $x$-axis into two equal halves, forming training and validation sets containing 1.13 million and 1.14 million galaxies, respectively. Each subset spans a volume of $(0.5~\mathrm{Gpc}) \times 1~\mathrm{Gpc} \times 1~\mathrm{Gpc}$. We adopt an equal split to ensure a large validation set, which supports an effective model evaluation even with low signal-to-noise data, and lets us apply interpretability techniques.

During training, the model predicts $\mu_i$ and $\kappa_i$ for the $i$th galaxy, which define a von Mises distribution over its orientation. We then maximize the joint log-likelihood of the observed angles under these distributions. We do this by minimizing the negative log-likelihood loss, averaged over the $N$ galaxies in the dataset,
\begin{equation}
    \mathcal{L}_\text{NLL} = -\frac{1}{N}\sum_{i=1}^N \log P_\text{VM}(\theta_i; \mu, \kappa),
\end{equation}
where $\theta_i$ is the $i$th galaxy's target orientation, and $P_\text{VM}(\theta_i; \mu_i, \kappa_i)$ is given by Equation~\ref{eq:von_mises}. For each model, we pre-train the GNN model for 2000 epochs (that is, 2000 complete passes through the training set), and train the complete probabilistic model for a further 2000 epochs by minimizing $\mathcal{L}_\text{NLL}$. 

We find that pre-training the GNN component before integrating it into the full probabilistic model improves convergence. During pre-training, we minimize the mean squared error between the GNN’s first equivariant latent vector and the target orientation. Since this objective closely resembles the task of predicting $\mu$, the first latent vector learns to encode much of the information relevant to the mean orientation. The remaining latent vectors are unused during this stage. Once we enable the full probabilistic model for standard training, these additional latent vectors allow the model to learn $\mu$ and $\kappa$ more effectively.

We find that passing the entire training dataset as a single batch yields a lower final loss compared to standard mini-batch training (e.g., batches of 128, 256, or 512 galaxies). We hypothesize that this improvement results from reduced gradient variance due to averaging over noisy labels, although we have not yet conducted a systematic study of batch size scaling. To reintroduce stochasticity into the optimization process, we randomly mask half of the galaxies' contributions during the loss aggregation step. 

However, processing over a million galaxies in a single batch requires substantial GPU memory. To make the computation tractable for this early demonstration of DELTA, we restrict all fully connected layers in DELTA -- $f_0$, $f_h$, $f_e$, and $f_w$ -- to a single hidden layer with 16 nodes and ReLU activation. We also fix the dimensionality of the node features, edge features, and latent vectors to 16. We anticipate increasing model size in future iterations will improve performance, especially with full-scale cosmology datasets.

\subsubsection{Model Evaluation Metrics}

DELTA’s training procedure differs from standard approaches in that the training objective is to match the distribution of noisy orientations in the data, while the scientific goal is to recover their mean. Recovering their mean isolates the pure, noise-free, tidally induced orientation signal. In observational data, no ground truth exists for the underlying tidal orientation. Instead, we assess model performance by comparing its predictions to a random baseline: if the model consistently outperforms random guessing across the noisy dataset, we infer that it has successfully identified information related to intrinsic alignments. When using mock datasets we also have the opportunity to directly compare model predictions against a known, noise-free tidal ground truth, providing a clearer benchmark for evaluating the model’s ability to recover intrinsic alignment signals.

Given we have access to these ground truths, we use two metrics to evaluate DELTA's angular prediction error: one relative to the noisy alignments (emulating observational data) and one relative to the pure alignments (verifying the model capabilities). Both metrics rely on an angular version of the mean absolute error, $e_\theta$, which accounts for the circular nature of angles and treats clockwise and counterclockwise errors equivalently. Given a set of target angles $\alpha_i$ and predicted angles $\beta_i$, each of length $N$, the error is defined as
\begin{equation}\label{eq:circular_error}
    e_\theta(\alpha, \beta) = \frac{1}{N} \sum_{i=1}^{N} \operatorname{mod}(|\beta_i - \alpha_i|, \pi).
\end{equation}

To make this metric clear, we quote it as an improvement percentage over the error from a uniform random guess. This emphasizes how much predictive information the conditioning contributes, and therefore how much information DELTA has captured from the neighborhood-tidal ellipticity relationship. For two independent samples from a uniform distribution $r \sim U[0, 2\pi]$, the expected error is $e_\theta(r, r) = \pi/2$.

We define the improvement percentage for the noisy alignments as
\begin{equation}
    \mathcal{I}_N = \left( 1 - \frac{e_\theta(\phi, \mu)}{\pi/2} \right) \times 100\%,
\end{equation}
where $\phi$ denotes the noisy orientations and $\mu$ denotes the model's predicted mean orientations. This metric reflects the scenario encountered in observational data and captures both the model’s predictive accuracy and the inherent noise in the alignment targets.

Similarly, we define the improvement percentage for the pure alignments as
\begin{equation}
    \mathcal{I}_P = \left( 1 - \frac{e_\theta(\hat\phi, \mu)}{\pi/2} \right) \times 100\%,
\end{equation}
where $\hat\phi$ represents the noise-free, tidally induced orientations. This metric indicates how effectively the model captures the underlying, noise-free intrinsic alignments injected into the data. In contrast to the noisy metric, it isolates the model's capability by removing the influence of observational noise. While this improvement cannot be measured in practice with real data, it offers valuable insight into model performance in the controlled setting of the mock demonstration. 

For both metrics, we report a 95\% confidence interval on the improvement percentage, obtained by bootstrapping over the $e_\theta$ mean computation. An improvement percentage of 100\% indicates that the model perfectly recovers the orientations. A value of 0\% implies that the model has not extracted any information from the neighborhood to inform its predictions, and is equivalent to a random guess. Negative values may occur if the model performs worse than a random guess as a consequence of overfitting, for example when collapsing onto predictions along a single axis.


\subsubsection{Results}

DELTA successfully recovers tidal alignments from noisy galaxy orientation data in mock datasets for all alignment scenarios. Table~\ref{tab:results} reports the improvement percentages over a random baseline for both the pure and noisy alignment metrics. In all three alignment scenarios, the model successfully captures a portion of the underlying satellite and central galaxy alignments. 

In both the medium and high alignment scenarios, DELTA recovers satellite alignments almost as well as the full alignment scenario, and actually improves upon the full scenario for central alignments. This is an indication that the symmetries of the EGNN allow it to effectively overcome the noise and perform close to as well as it would if there was no noise. 

There is a sharp decrease in performance between medium and low alignment scenario, suggesting a non-linear scaling between alignment strength and recovered alignments. Physically, this indicates DELTA will exhibit optimal performance for intermediate- to high-mass galaxies. 

\begin{figure*}
    \centering
    \includegraphics[width=\linewidth]{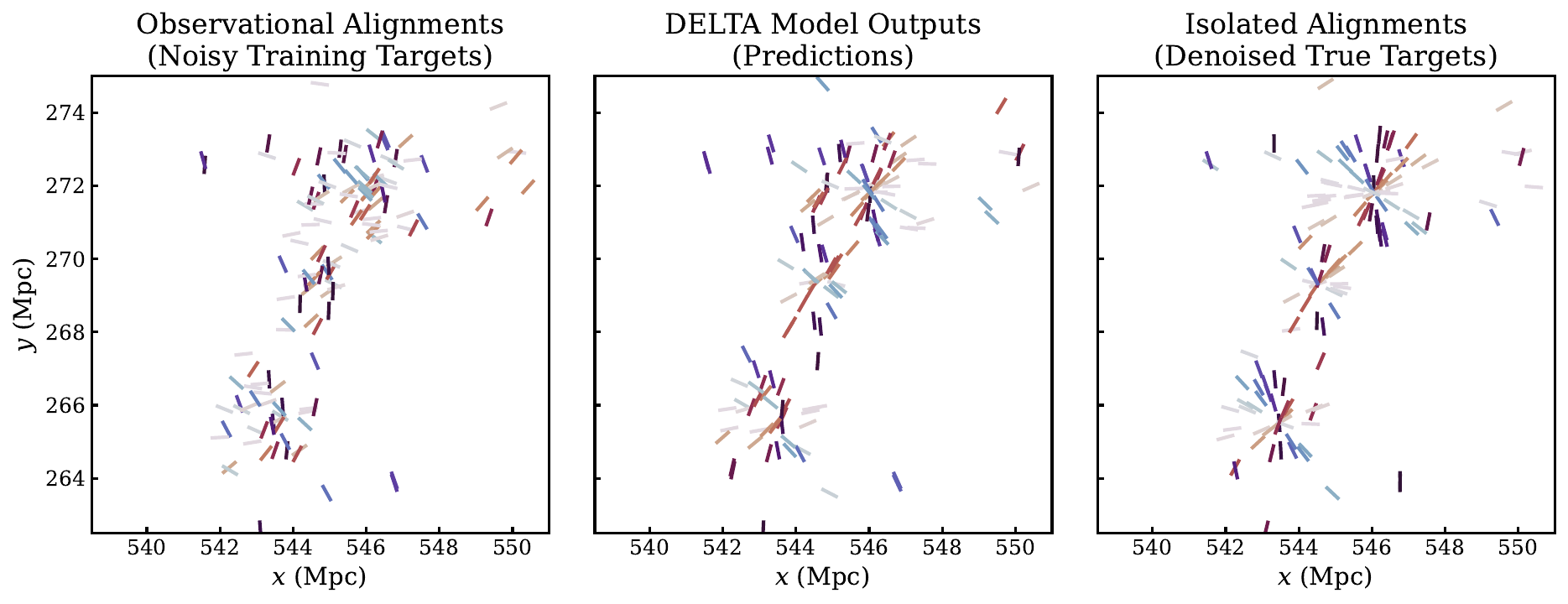}
    \caption{Mapped intrinsic alignments for a region of the validation dataset, demonstrating DELTA’s ability to recover intrinsic alignments from noisy data. Each line segment's angle represents the orientation of its corresponding galaxy, with colors by angle for visual guidance. \textit{Left:} Noisy intrinsic alignments for the high alignment scenario. These noisy alignments combine the alignment signal with realistic random orientation noise. These galaxy orientations are comparable to observational data. These noisy orientations act as the training targets in the DELTA model. \textit{Middle:} Galaxy alignments predicted by DELTA, using only the positions of surrounding galaxies as inputs and noisy orientations (left panel) as targets. These are the model's learned approximation of the underlying de-noised alignment signature, reconstructed by learning the average patterns across the training data. \textit{Right:} Pure noise-free alignments from the simulation, representing a perfect recovery of alignments. DELTA makes a clear partial reconstruction of the radial alignments for satellite galaxies.}
    \label{fig:alignment_map}
\end{figure*}

To visualize DELTA’s performance, Figure~\ref{fig:alignment_map} shows the recovered intrinsic alignments compared to both the noisy and pure alignments for the high alignment scenario. This visualization shows that the learned $\mu$ of the von Mises distribution captures much of the pure radial alignment signature of satellite galaxies, despite only seeing noisy variations of this signature throughout training. This setup allows DELTA to cut through the noise and capture the pure alignment signal that one would expect for every galaxy. Visually, it is clear that the reconstruction is not perfect, but DELTA definitively identifies structure that is completely invisible in the noisy training targets alone. 

In general, DELTA captures satellite alignments more strongly than central alignments, likely due to their difference in complexity. While satellite galaxies point towards cluster centers (which are well sampled with many galaxies), central galaxies instead align with the halo alignments directly extracted from the dark matter clouds from N-body simulations. As a result, the central alignments are a much more complex signature, dependent on non-linear gravitational interactions. However, DELTA's nonzero recovery of this signature is promising, especially given the size limit imposed on DELTA's EGNN model. Extending this to a larger EGNN would increase the informational capacity and allow DELTA to capture more of the complex central alignment signature. 
 
To evaluate DELTA’s applicability to observational data, where pure alignments are not available, we also compute the improvement percentage relative to the noisy alignment targets. In all three alignment scenarios, we identify a nonzero improvement percentage. These results confirm that DELTA can successfully learn tidally-induced alignment in galaxies and verify their presence without requiring access to noise-free labels, supporting its use in real survey data where only noisy observations are available.

We note that $\mathcal{I}_N = 100\%$ is not achievable, even under perfect model performance, due to the inherent noise in the targets. The maximum possible values of $\mathcal{I}_N$ are 7.7\%, 13.9\% and 22.8\% for the low, medium and high alignment scenarios respectively. We obtain these values by computing the error of the noisy orientations $\phi$ relative to the pure orientations $\hat\phi$ using $e_\theta(\phi, \hat\phi)$. These bounds reflect the limit imposed by the noise in the evaluation targets.

\begin{table*}
    \centering
    \renewcommand{\arraystretch}{1.2}
    \begin{tabular}{|l|c|c|c|c|c|c|c|c|}
        \hline
        \multirow{2}{*}{\textbf{Scenario}} & \multicolumn{2}{c|}{\textbf{Alignment Strength}} & \multicolumn{3}{c|}{\textbf{Noisy Improvement} $\mathcal{I}_N$ (\%)} & \multicolumn{3}{c|}{\textbf{Pure Improvement} $\mathcal{I}_P$ (\%)} \\
        \cline{2-9}
        & Central & Satellite & Central & Satellite & Total & Central & Satellite & Total \\
        \hline
        Low & 0.55 & 0.10 & 1.4 & 0.4 & 1.0 & 8.5 & 15.6 & 11.3 \\
        Medium & 0.70 & 0.25 & 2.9 & 2.9 & 2.9 & 10.9 & 65.7 & 32.1 \\
        High & 0.80 & 0.45 & 4.7 & 6.9 & 5.6 & 11.0 & 69.9 & 33.7 \\
        Full & 1.00 & 1.00 & -- & -- & -- & 10.1 & 82.2 & 37.9 \\
        \hline
    \end{tabular}
    \caption{Percentage improvement over a random guess for predicted intrinsic alignments, shown for varying intrinsic alignment scenarios. The low, medium and high alignment scenarios correspond to plausible alignment noise levels according to hydrodynamics simulations, for lower mass, intermediate mass, and high mass galaxies respectively. The full alignment scenario aligns galaxies perfectly according to the alignment model, with no added directional noise. All improvement values include a 95\% confidence interval of $\pm0.1\%$, estimated via bootstrapping over the validation dataset. The noisy improvement metric $\mathcal{I}_N$ evaluates performance relative to the noisy alignment targets used during training, which include both intrinsic alignments and stochastic shape noise (akin to the improvement we would see in observational data). The pure improvement metric $\mathcal{I}_P$ compares predictions to the de-noised, ground-truth alignments, only available in the simulation. Recovering the pure alignments ($\mathcal{I}_P=100\%$) is the goal of the DELTA model. We show no noisy improvement for the full alignment scenario, since it has no noise and trains directly on the pure alignments.}
    \label{tab:results}
\end{table*}


\begin{figure}
    \centering
    \includegraphics[width=\linewidth]{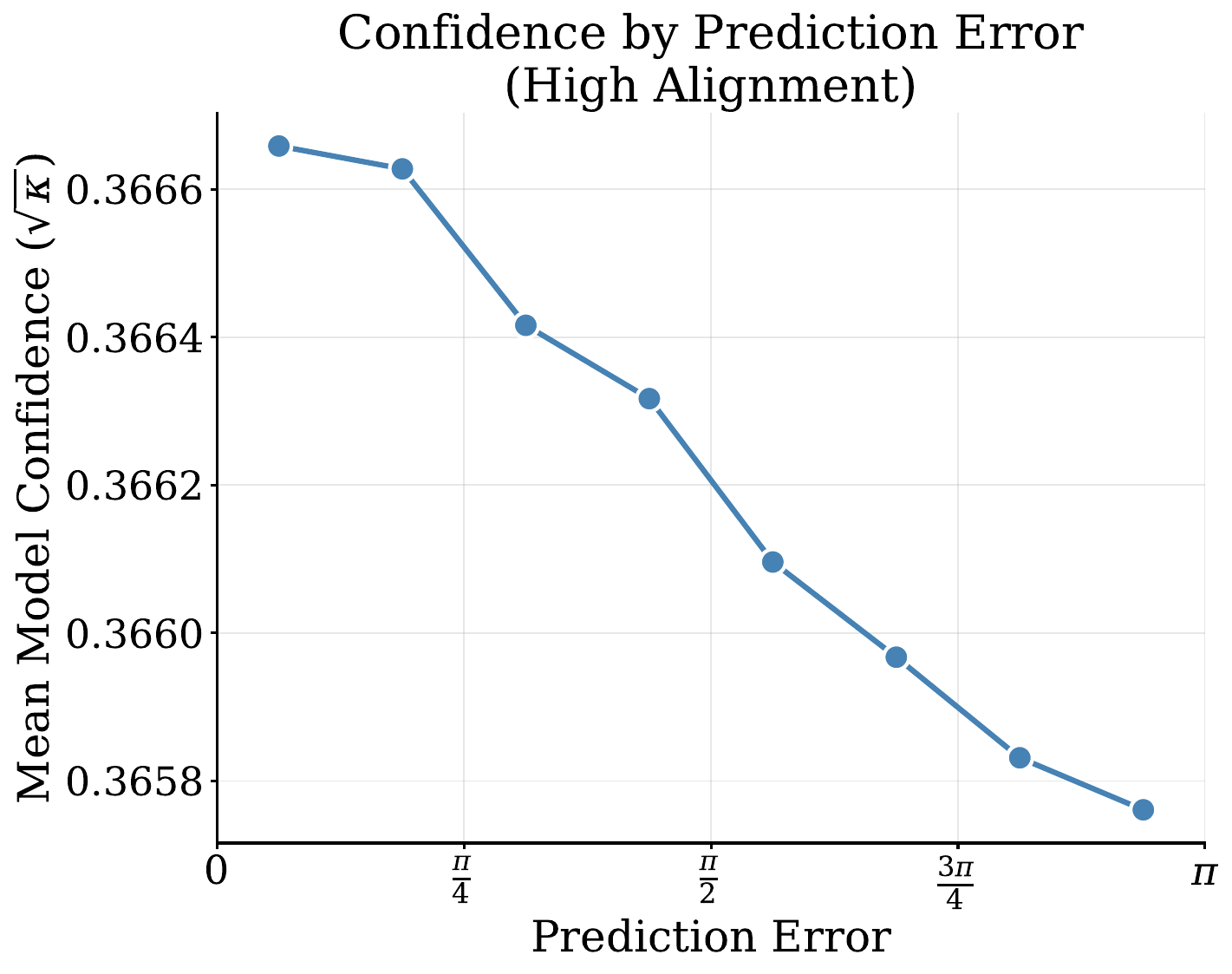}
    \caption{Mean prediction confidence, binned by prediction error, for intrinsic alignments predicted by DELTA. An inverse correlation between error and confidence indicates that the model learns to identify when its predictions are less reliable.}
    \label{fig:error_confidence_relation}
\end{figure}

To illustrate the effectiveness of the probabilistic modeling approach, Figure~\ref{fig:error_confidence_relation} shows the relationship between prediction accuracy and confidence for the high alignment scenario. We define prediction confidence as $\sqrt{\kappa}$, which serves as an analogue to the inverse standard deviation of the von Mises distribution. For each galaxy, we compute its prediction confidence and group it into bins based on the corresponding prediction error relative to the noisy alignments, $e_\theta(\phi, \mu)$.

We observe a clear inverse trend: predictions with higher accuracy tend to exhibit higher confidence, while less accurate predictions correspond to lower confidence. This relationship indicates that the model has learned to recognize when its predictions are less reliable and adjusts its confidence accordingly. When such a relationship holds, it enables us to distinguish confidently predicted alignments from uncertain ones, a property that is useful for interpreting the model's predictions from a physical perspective, and also valuable for understanding uncertainties in downstream tasks.

\subsection{Discussion} \label{sec:delta_discussion}

We have demonstrated DELTA’s ability to recover intrinsic alignments in mock catalogs using only noisy data, analogous to that available in observational surveys. The alignment maps produced by DELTA clearly reveal the expected intrinsic alignment structure in a readily interpretable form. Even without prior knowledge of the injected alignment model, one could infer from these maps that satellite galaxies exhibit strong radial alignment with their cluster centers, suggesting the presence of an underlying physical mechanism. 

This result not only shows that DELTA can extract alignment signals from data dominated by isotropic shape noise, but also suggests that it could similarly disentangle alignments from lensing-induced distortions, which appear isotropic to the model in the absence of line-of-sight information. Overall, DELTA shows strong potential as a method for accurately modeling intrinsic alignments, disentangling tidal and lensing contributions, and offering new insight into the physics of galaxy alignments.



\subsubsection{Leveraging Deep Learning Directly on Data}

DELTA combines the flexibility of deep learning with a fully data-driven approach, avoiding domain shift errors associated with models that train on simulated large-scale structure and apply to observational data. Deep learning's high representational capacity can lead to overfitting to features that are unique to the training simulation, raising concerns about the model's inferences on observational data. These domain shift errors can be controlled by limiting model complexity (e.g \cite{regaldosaintblancard2024, gatti2025}), but this comes at the cost of limiting model expressiveness. 

In contrast, DELTA avoids this tradeoff, because both training and validation occur on data from the same observational pipeline, mitigating the risk of domain shift between training and application data. This setup allows us to retain greater model complexity while also maintaining robustness. Moreover, by validating the model on held-out data, we can directly verify the presence of learned alignment signals in the observations themselves. Nonetheless, care must be taken to distinguish true physical signals from systematic effects, a topic we address in Section~\ref{sec:systematics}.

This added model complexity is particularly valuable for modeling intrinsic alignments. DELTA’s flexible, data-driven framework enables it to accommodate modeling uncertainties that affect traditional approaches. One example is the mismatch between the observed galaxy distribution and the underlying dark matter field, known as galaxy bias. Because tidal forces are sourced predominantly by dark matter, this bias complicates intrinsic alignment modeling, which relies on visible galaxies. DELTA's flexibility means it can learn an internal mapping from galaxy positions to tidal ellipticity contribution (which depends on the full matter distribution), effectively constructing an implicit model of galaxy bias.

The same flexibility allows DELTA to account for other complex, higher-order effects that are difficult to capture with parametric models, including nonlinear tidal interactions, environmental dependencies, and baryonic influences on galaxy orientation. DELTA also avoids several of the limitations faced by current IA models, such as incorrect redshift scaling and poor performance on small scales \citep{georgiou2019, johnston2019, desy3shear, chen2024}.

Our demonstration further supports this capability: the injected alignments are defined with respect to dark matter halo orientations rather than the galaxy distribution, yet DELTA successfully recovers them. This suggests that the model has implicitly learned both aspects of the galaxy-halo connection and the alignment relationship relative to halos.

\subsubsection{Avoiding Contamination from Lensing}

To isolate the intrinsic alignment contribution, we assume that the lensing signal is independent of the local neighborhood, such that $P(\epsilon_L | \mathcal{N}) = P(\epsilon_L)$. Under the assumption of a statistically isotropic universe, the lensing contribution averages to zero, and the model cannot predict any lensing-induced ellipticity. 

To ensure that the model does not inadvertently learn the lensing signal, the neighborhood must exclude any information that correlates with lensing. Otherwise, the model could use this information to infer lensing-induced shapes, contaminating the tidal alignment signal. In particular, this excludes galaxy properties that are directly influenced by lensing, such as peak luminosity or apparent size. Gravitational lensing is well understood through general relativity, and careful selection of galaxy properties can minimize the risk of lensing contamination. Several ratio-based quantities remain unaffected by lensing and can be safely used as input features:
\begin{itemize}
    \item \textit{Galaxy Color}: Lensing magnification affects all flux bands equally. Because color is a flux ratio, it remains invariant under lensing. This invariance extends to derived quantities such as star formation rate, stellar temperature, and dust reddening.
    \item \textit{Galaxy Surface Brightness}: Lensing magnification increases both the observed flux and the apparent area by the same factor, leaving the surface brightness defined by their ratio unchanged.
    \item \textit{Size Ratios}: Many morphological parameters rely on ratios of galaxy sizes. Since both components scale equally under magnification, these ratios remain unaffected. Examples include the concentration index and the bulge-to-disk ratio.
\end{itemize}

DELTA also explicitly restricts the GNN to use only galaxy position information as input to predict orientation. Information about galaxy orientation is never used as an input, or shared between nodes. As a result, the model cannot directly learn correlations between galaxy shapes, only the correlations between positions and shapes. This design mirrors previous intrinsic alignment analyses that rely on shape-position correlations while excluding shape-shape correlations \citep[e.g.][]{hirata2007, singh2015, johnston2019}, and extends this principle to the broader set of features that the deep learning model can extract.

A related point is that we must use spatially distinct training and validation sets. If training and validation data are spatially intermixed, the model could potentially learn to associate a specific background configuration with a foreground lensing signal, which would also appear in the validation set. In contrast, if the validation data occupies an entirely separate spatial region, such correlations cannot transfer and would not contribute to a successful model. Although overfitting to specific background-foreground configurations may occur within the training data, the large volume and sample size of cosmological surveys reduce the likelihood that the model will favor localized pattern matching over globally consistent relationships.

In the context of real data, we can assess whether such overfitting occurs by comparing performance on the training and validation sets. A significant discrepancy would indicate that the model is exploiting spurious correlations, while comparable performance supports the interpretation that DELTA is learning robust, generalizable patterns in the data.

However, despite these controls, some limitations remain in the assumption that lensing is independent of the neighborhood.

Firstly, redshift uncertainties and redshift-space distortions cause some mixing of relative positions along the line-of-sight. Misplaced galaxies could in principle introduce lensing-related information into the neighborhood. However, such cases represent only a small fraction of the galaxies that contribute to the lensing component $\epsilon_L$. Moreover, the lensing efficiency is lower for matter close to the source galaxy, further suppressing the contribution from these misplaced foreground galaxies. As a result, DELTA is unlikely to learn any significant lensing-induced shape from these galaxies. However, redshift uncertainties do introduce noise in the line-of-sight position. We discuss the implications of this line-of-sight noise in greater detail in Section~\ref{sec:systematics}.

Secondly, gravitational lensing can cause small deflections in the apparent positions of galaxies, shifting them slightly from their true locations. However, this effect is subtle, and any lensing information that the model might extract from such positional shifts is expected to be subdominant compared to the much stronger relationship between the local environment and intrinsic alignments. 

Thirdly, some correlation between lensing galaxies and neighborhood galaxies is inevitable due to the underlying coherence of large-scale structure. However, these correlations are again expected to be subdominant.

In future work, we will further investigate the assumption that lensing is independent of the neighborhood to ensure that any such correlations do not significantly affect the results.

\section{Intrinsic Alignment Insights with Deep Learning Interpretation}\label{sec:interpretation}
DELTA is not only a powerful tool for recovering intrinsic alignments, but also provides insight into the underlying alignment physics. The most basic insight comes from a direct visualization of alignment maps, but we can extend our insights with specialized deep learning interpretability techniques. Due to the complexity of DELTA’s EGNN backbone, it is challenging to directly understand what it has learned about the physics of tidal interactions. 

Interpretability methods help rank the importance of input features, understand internal processing, and uncover nontrivial dependencies that may not be apparent from direct inspection. By applying these methods, we can characterize intrinsic alignments in unprecedented detail, advancing our understanding of their physical origins and informing the development of more accurate IA models. In doing so, we also gain new perspectives on the processes governing galaxy evolution.

In this section, we demonstrate two GNN interpretability techniques designed to reveal how different components of the input dataset contribute to the model’s predictions. The first, a permutation test \citep{breiman2001}, assesses the contribution of galaxy properties to the predicted tidal alignments. The second, a SHAP test \citep{lundberg2017}, quantifies the individual influence of each neighboring galaxy on a given galaxy’s predicted alignment. Given our precise knowledge of the injected IA signals in the data, we can qualitatively compare the interpretability-driven characterization to the injected model. This comparison allows us to assess whether the interpretation techniques effectively recover the expected alignment features. 


\subsection{Permutation Test for Galaxy Observables}\label{sec:permutation_test}
Many galaxy observable properties are inherently linked with the surrounding large-scale structure and tidal environment and by extension, intrinsic alignments. For example, galaxy color is influenced by star formation history, which is in turn dependent on the density of gas and dark matter in the galaxy's intergalactic environment, mergers between galaxies, and tidal interactions. As a result, galaxy observables can provide additional information to more accurately predict intrinsic alignments. Such an effect is already widely used in intrinsic alignment analysis in the TATT model, in which alignments depend on galaxy morphology, which is itself determined through galaxy observable properties such as color and profile. 

Permutation tests provide a simple method to assess the influence of different galaxy observables on DELTA's alignment predictions. When applied to GNN-type models the permutation test involves shuffling a specific node feature, such that each node now has a randomly assigned value of that property, and assessing the change in performance due to the shuffle. This randomization breaks any learned association between that property and the output, so a reduced performance implies that the node feature was used by the GNN and informative to the output.   

For DELTA, this means shuffling an observable property across galaxies, and assessing how the model responds. If the prediction error decreases, DELTA is using that observable property to inform alignment predictions, implying it has some nonzero relationship with intrinsic alignments. The permutation test's primary utility is its ability to identify a relationship of arbitrary complexity between model inputs and outputs, limited only by the expressive power of the deep learning model. While the test does not reveal the form of the relationship, it is computationally efficient and can be applied broadly across all observable galaxy properties, making it an excellent first step to understanding connected properties. 

For a dataset consisting of nodes with $M$ associated properties, we define the node properties as $\vec{p} = (p^0, p^1, p^2, \dots, p^M)$, where each $p^m$ represents the $m$-th property of a given node. The permutation test proceeds as follows:
\begin{itemize}
    \item Train the DELTA model using all $M+1$ node properties. The first property, $p^0$, is a control consisting of randomly assigned noise, while $p^m$ for $m \geq 1$ represent the true node properties.
    
    \item Use the trained model to predict the baseline tidal orientations, denoted $\mu$.

    \item For each node property $m \in \{0, 1, \dots, M\}$, shuffle $p^m$ across the dataset to break its association with the target, then use the model to produce new predictions $\mu^m_\text{perm}$.

    \item Measure the change in performance due to the permutation by computing
    \begin{equation}
        \Delta^m = e_\theta(\phi, \mu^m_\text{perm}) - e_\theta(\phi, \mu),
    \end{equation}
    where $\phi$ is the set of observed orientations and $e_\theta$ is the angular error defined in Equation~\ref{eq:circular_error}.
\end{itemize}

We repeat this procedure multiple times to account for stochastic variation in training outcomes. The model may occasionally converge on solutions that do not utilize specific input properties, even if they are useful. Repeated trials help ensure an informative property is not missed by chance. Informative properties are visually apparent by eye across the ten runs, but a more statistically rigorous approach would be suitable in future applications to data.

When applying to observational data, we would use the permutation test for observational galaxy properties such as color and profile. However, the mocks do not have any direct observable galaxy properties, so we instead use several properties derived from the simulation. For each galaxy, we include 5 properties: the galaxy's parent halo's log mass and virial radius, the distance between the galaxy and the halo center, a binary property representing the galaxy type (satellite or central), and the log local density in the galaxy's neighborhood. We assign the local density as the count of galaxies within a 5 Mpc radius. Before training, we standardize each galaxy property to zero mean and unit standard deviation across the dataset. 

\begin{figure}
    \centering
    \includegraphics[width=\linewidth]{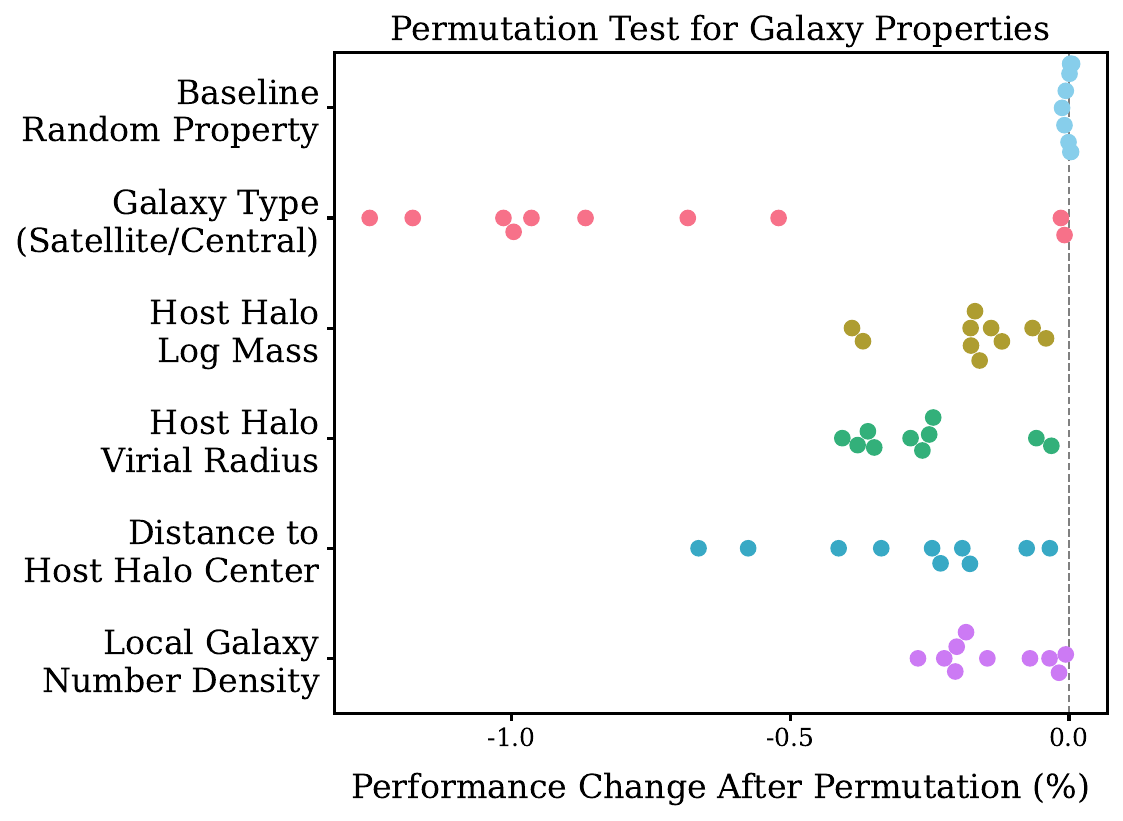}
    \caption{A permutation test for several galaxy properties, identifying the presence of relationships between properties and intrinsic alignments. The permutation test measures the decrease in DELTA's performance over the test set after shuffling the respective node property. A decrease indicates that that the property is informative to DELTA's predictions, and is therefore correlated with intrinsic alignments. The baseline random property shows no correlation with alignments, while all other properties show evidence of correlation.}
    \label{fig:observables}
\end{figure}

The permutation tests successfully identify correlations between the assigned properties and intrinsic alignments. To illustrate this detection, Figure~\ref{fig:observables} shows the change in percentage improvement $\mathcal{I}_N$ after permuting the baseline random observable property and the local number density property across ten full training and validation runs. Permuting the baseline random property results in no systematic change in performance, aside from some minor scatter due to model stochasticity. In contrast, permuting each of the other five galaxy properties leads to a visually apparent degradation in model performance.

The largest drop in performance, and therefore most influential property, is the galaxy type. This is expected given that it controls the type of injected alignment between radial or halo alignment, and is therefore highly beneficial for predicting galaxy orientation. The remaining four properties also all have a relationship with the intrinsic alignments.  

These connections all imply a physical link between the galaxy (or halo) property and intrinsic alignment physics (or in this case, the injected model). For example, the permutation test identifies a connection between local density and galaxy orientation. This matches the injected alignments: satellite galaxies statistically dominate high number density regions because these regions are populated by massive halos that each host many satellites, while central galaxies are statistically more numerous in lower density regions where smaller halos (typically hosting only one central galaxy each) are more common. In observational data, we will have access to reconstructed host halo properties and a wide array of direct observable properties, such as galaxy color and relative sizes, which provide a rich set of potential connections to probe.


\subsection{Latent Space Visualization}

Another approach to understanding a deep learning model's internal processing is to extract insights by visualizing the inner stages of the model. In the case of DELTA, the EGNN outputs an equivariant latent space, which then passes through some more non-linear processing to predict mean galaxy alignment direction and confidence. By studying how the magnitudes of these latent vectors correlate with the physical structure, we can identify how the model uses its surroundings to make predictions. 

In DELTA, we use a latent space corresponding to 16 latent vectors. After taking each vector's magnitude, the result is a 16 dimensional latent space, which is difficult to visualize and interpret. 

To make the space more visually accessible, we project it into a 2D representation using a UMAP (Uniform Manifold Approximation and Projection) dimensional reduction. UMAP is a non-linear dimensionality reduction technique that captures local and global structures in high-dimensional data \citep{mcinnes2020}. As a result, nearby locations in this UMAP projection are also nearby in the full high-dimensional space. We use UMAP to project the latent space magnitudes into a 2D embedding, using the default settings in the \texttt{umap-learn} python package \citep{mcinnes2020}. 

\begin{figure}
    \centering
    \includegraphics[width=1.0\linewidth]{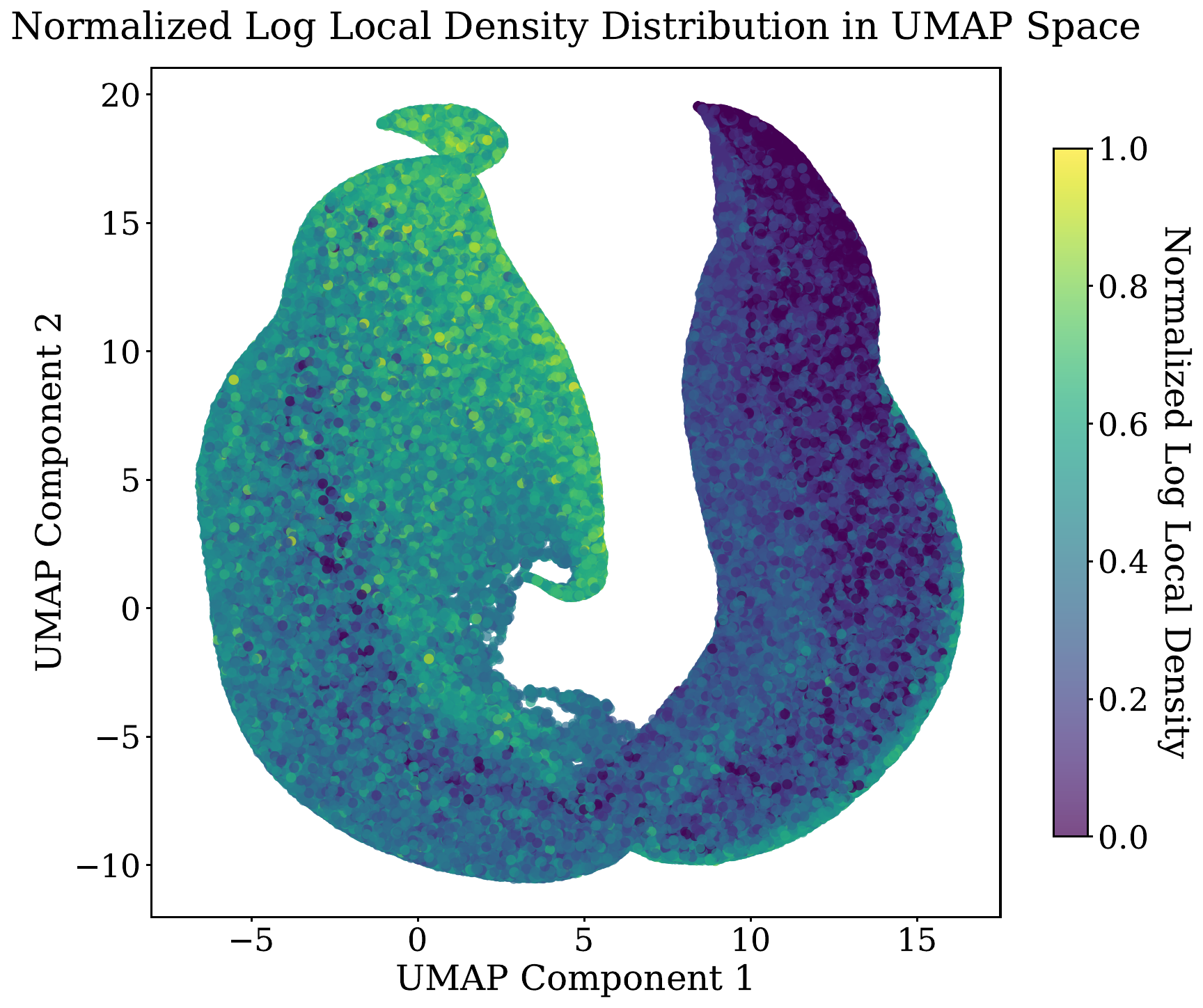}
    \caption{A UMAP projection of the magnitude of latent space vectors learned by the equivariant graph neural network. Each point represents a galaxy, and is colored by normalized log density as determined by the number of nearby galaxies within a 5 Mpc sphere. The smooth change in density implies that galaxies with similar densities experience similar alignments mechanisms. Structures in this UMAP projection can reveal differing modes of galaxy alignments.}
    \label{fig:umap_density}
\end{figure}

In Figure~\ref{fig:umap_density} we plot the learned 2D projection and color it by each galaxy's local density. This plot reveals that the EGNN's latent space smoothly orders galaxies by density, indicating that density is an important feature when predicting alignments. Going further, the distant spatial separation of low and high density galaxies shows that they exist in very different regions in the latent space. This implies that denser regions (clusters) have different intrinsic alignments than those in sparser regions (voids and filaments), suggesting distinct physical processes driving the alignment in these regions. In this case, rather than physical processes, we are capturing the difference between alignment models for satellite and central galaxies in the mocks, where satellite galaxies are the dominant type found in dense regions. 

\begin{figure}
    \centering
    \includegraphics[width=1.0\linewidth]{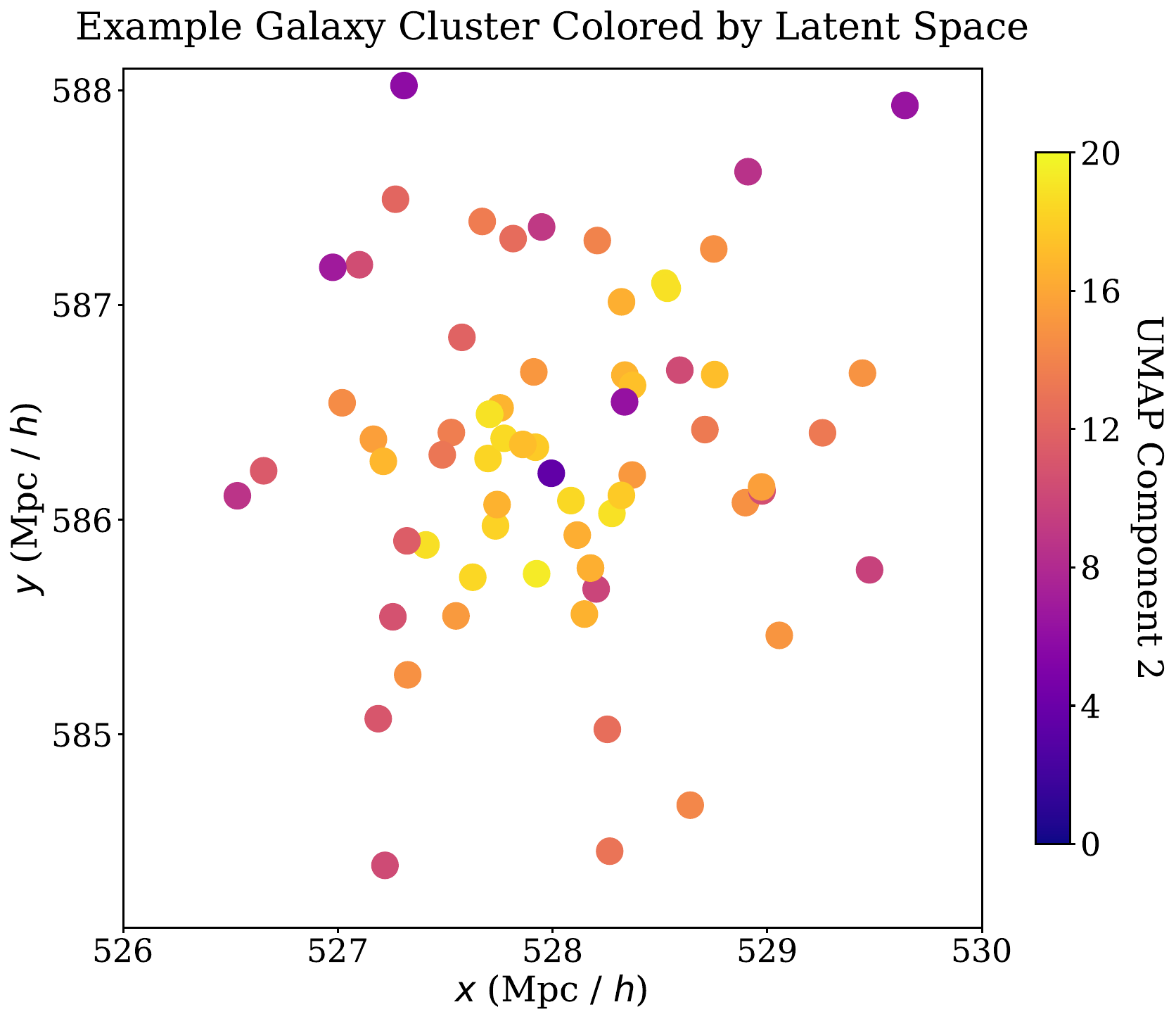}
    \caption{An example galaxy cluster colored by the second UMAP component corresponding to the $y$-axis in Figure~\ref{fig:umap_density}. The central galaxy has a clearly distinct low value compared to the high values of its neighbors. This is an indication that central galaxies align differently to their nearby satellite galaxies. This reflects the true nature of the assigned alignments, demonstrating how DELTA can both learn and identify non-trivial alignment features.}
    \label{fig:umap_cluster}
\end{figure}

Studying the UMAP space further, we notice that for high density cluster regions, there are equally-dense regions with significant separation in UMAP space, along UMAP Component 2. This indicates some further variation in alignment physics between different galaxies within clusters. To investigate effect, in Figure~\ref{fig:umap_cluster} we plot a galaxy cluster, colored by UMAP Component 2. This distribution reveals a clear pattern: galaxies within the cluster's center tend to experience different alignments to those in the outer regions of the cluster. This implies that central galaxies experience different alignment physics to satellite galaxies. Again, this conclusion accurately reflects the injected alignment signal.

Through this visual analysis of latent space clustering, we can gain powerful insights into how the model processes physical features. We could apply both analysis techniques described here directly to observational data, since they do not rely on any ground truths from simulations that are unreachable in data. In an application to data, this interpretability technique can reveal new patterns and lead to new insights into alignment physics.

These interpretability results demonstrate that DELTA learns genuine physical relationships rather than spurious correlations, addressing concerns about deploying deep learning in cosmological contexts. The ability to validate learned alignments through interpretability techniques will be particularly valuable when applying DELTA to observational data, where ground truth alignments are unavailable for direct comparison.

\subsection{Discussion}

We demonstrate how interpretability techniques provide insight into the DELTA model and in turn reveal the physics driving intrinsic alignments. The typical approach to understanding alignments requires formulating specific functional forms from prior assumptions about the alignments, and constraining these in conjunction with a full weak lensing analysis. This new approach is instead free-form, using deep learning to capture arbitrary physical relationships and interpretability to reveal these relationships.  

The permutation test is a simple but powerful interpretability approach that, when combined with DELTA, identifies correlations between galaxy observables and intrinsic alignments. Unlike traditional approaches that test for specific correlations that are assumed from prior knowledge, this method is model-agnostic and can discover unexpected connections. We demonstrate the detection of correlations between several simulated properties, accurately reflecting expectations from the injected alignment signal. Among these is a detected correlation between local galaxy density and alignments, which can be directly applied to observational data. 

Latent space visualization is a complementary interpretability approach that provides more detailed insights into how DELTA processes spatial information. It assesses how the EGNN compression model groups galaxies in high-dimensional space, providing insights into how galaxies in different environments are assigned alignments. This approach can uncover emergent patterns that may not be apparent with traditional correlation analyses. In the mock dataset, we demonstrate the identification of a density-dependence and two separate alignment modes in high-density regions, both of which accurately reflect the injected alignment signal.

These interpretability methods help address concerns about deploying deep learning in scientific contexts. By better understanding DELTA's processing, we can be confident that it has learned physically meaningful correlations in the data rather than artifacts and systematics. This validation is important in intrinsic alignment studies, where systematics can bias final cosmology results. Furthermore, the integration of interpretability techniques with deep learning provides a new avenue for scientific discovery. We can use deep learning models to capture complex physical patterns in data that would otherwise go un-noticed, and use interpretability techniques to extract these patterns from the model. 

\section{Future Application to Observational Data}\label{sec:discussion}
In this work, we demonstrate the success of DELTA on simulated mock catalogs with nontrivial injected intrinsic alignments. However, the ultimate objective is to apply DELTA to observational survey data, enabling the characterization of intrinsic alignments in real galaxies. In this section we discuss the requirements that the observational dataset will need to satisfy and the complications from systematic biases.

\subsection{Observational Dataset Requirements}

To effectively isolate galaxy intrinsic alignments, DELTA requires a specialized dataset with two core components across a large catalog of galaxies: shape measurements and 3D galaxy positions. Shape measurements are readily available in next-generation photometric surveys. In the coming years, the Euclid Wide Survey and Rubin's Legacy Survey of Space and Time will provide shape measurements for billions of galaxies across more than half the sky \citep{euclid2024, rubin2019}. 

The primary limitation in applying DELTA to real data lies in obtaining accurate 3D positions, particularly along the line of sight. While photometric redshifts provide rough estimates of line-of-sight positions, they are too imprecise to reliably identify a galaxy’s neighborhood. Instead, DELTA requires the precision of spectroscopic redshifts.

A strong option is to combine 3D positions from the Dark Energy Spectroscopic Instrument (DESI) survey \citep{desi2022} with shape measurements from Euclid. These surveys have a $\sim$9000 square degree overlap \citep{naidoo2023}, yielding approximately 20 million galaxy shape and 3D position pairs, which is an order-of-magnitude increase over the mock catalogs used in this work. 

As an alternative to the high-resolution spectrograph of DESI, we may also consider grism-based redshift estimates. Euclid's NISP instrument will provide both grism redshifts and shape measurements for $\sim$50 million galaxies. Although grism redshifts are less precise, the larger dataset and uniformity of systematics across shape and position measurements may offset the reduced precision. We will explore the impact of reduced redshift precision in future work.

\subsection{Impacts of Observational Systematic Biases}\label{sec:systematics}
When measuring galaxy shapes, observable properties, and positions, various observational systematics can contaminate the data, leading to misrepresentations of the true galaxy properties, which is a common challenge across many cosmological analyses. DELTA’s approach, which involves training and validating on subsets of the same dataset that share the same systematics, offers a unique robustness to these effects. 

To consider the impacts of observational systematics on DELTA, we categorize them into two types: those that affect the model inputs (galaxy positions and observable properties), and those that affect the model output targets (galaxy orientations).

First, consider systematics that affect galaxy positions and observable properties. Examples of this systematic include survey incompleteness from survey masking and fiber collisions, and redshift errors due to mis-estimation and redshift space distortions. These systematics introduce additional complexity into the relationship between a galaxy’s neighborhood and its orientation, as the model must learn both the underlying physical connection and compensate for any systematic bias in order to make accurate predictions. However, provided these systematics appear equally in the training and validation, this is roughly equivalent to weakening the signal strength of the structure-alignment connection, which should be absorbed as additional noise in the probabilistic prediction. 



The second type of observational bias affects the measured galaxy orientations, which serve as targets during training. Systematic shifts in the targets influence what DELTA learns as the true, noise-free alignment signal, potentially skewing the predicted alignments toward observational biases rather than underlying physics. Significant effort has been devoted to controlling shape measurement biases in weak lensing analyses, particularly through shear calibration methods such as \textsc{METACALIBRATION} \citep{sheldon2017}. Using these calibrated shapes should reduce the impact of such uncertainties on the alignment model. 

In addition, we can further mitigate this type of systematic through careful application of interpretability techniques. By analyzing the features DELTA relies on to make its predictions, we can verify that the learned alignments reflect genuine physical relationships rather than artifacts introduced by observational systematics. The two methods we demonstrate are part of a broader class of deep learning interpretability tools designed to make complex models more transparent \citep{wu2022}. As this field continues to advance, we can apply more sophisticated interpretability techniques to DELTA to identify the presence of any systematics. 

In future work, we will conduct a more detailed investigation into the implications of specific systematics in recovered galaxy orientations and inferred intrinsic alignments. When applied to observational data, these approaches can reveal novel insights into alignment physics, leading to more accurate intrinsic alignments modeling and ultimately improved cosmological analyses.

\section{Conclusion}\label{sec:conclusions}
We introduce the Data-Empiric Learned Tidal Alignments (DELTA) model, a deep learning framework for isolating and characterizing galaxy intrinsic alignments directly from data. DELTA departs from traditional approaches by leveraging the principle that intrinsic alignments are correlated with the local galaxy distribution, while lensing distortions and stochastic shape noise are not. This allows the model to isolate the tidally-induced ellipticity contribution using only the observed properties and positions of nearby galaxies.

We design a specialized architecture for DELTA. It first uses an Equivariant Graph Neural Network (EGNN) for compressing local structure while preserving the fundamental symmetries of the large-scale structure. It combines this with a probabilistic output that models orientation uncertainty using a von Mises distribution. This architecture balances broad global assumptions with localized flexibility to capture the complex tidal physics driving intrinsic alignments, but in a way that handles the natural noise-dominated nature of galaxy orientations.  

We validate DELTA using mock datasets with injected intrinsic alignment signals. Despite being trained solely on noisy orientations, the expected value of the model's probabilistic predictions accurately recovers the underlying noise-free tidal alignment field. DELTA reduces the error on alignment predictions over a random guess by 8-11\% for central galaxies and 16-70\% for satellite galaxies, across a low, medium and high alignment scenarios. These alignment scenarios reflect observed alignments in realistic hydrodynamical simulations, so these results heavily imply that DELTA is capable of capturing alignments in observational data. We also demonstrate that alignment capture is identifiable even without a ground truth, a necessary requirement for application to observational data. 

To probe the model’s internal mechanisms and gain physical insight, we apply two deep learning interpretability techniques. A permutation test evaluates the relationships between different galaxy properties on alignment predictions, while a latent space analysis reveals how DELTA internally organizes and processes complex environmental information. These techniques allow us to connect the model’s outputs to physical drivers of alignment and demonstrate that it has learned meaningful relationships embedded in the data. Even without prior knowledge of the injected alignment model, these tools recover relevant learned features and show the role of local environment in shaping alignments. Both of these interpretability require only features that are available in observational data. 

We plan to characterize DELTA's performance under varying levels of signal-to-noise and survey volumes. We also aim to study the effects of observational systematics on recovered alignments and evaluate how DELTA absorbs or mitigates these uncertainties. The ultimate goal is to apply DELTA real observational survey data. A promising option is the combination of upcoming Euclid, Rubin and DESI datasets, which will provide shapes and positions of millions of suitable galaxies.

More broadly, DELTA represents a promising new approach for using deep learning in cosmological inference. Its ability to extract complex, nonparametric alignment signals from data without relying on fixed templates or simulations, and its compatibility with interpretability tools, makes it a powerful option for providing deeper insight into intrinsic alignments, galaxy evolution and broader cosmology.

\begin{acknowledgments}
M.C. acknowledges the support of an Australian Government Research Training Program (RTP) Scholarship. M.C., R.R., T.M.D., acknowledge the support of an Australian Research Council Australian Laureate Fellowship (FL180100168) funded by the Australian Government. Y.S.T is supported by the National Science Foundation under Grant No. AST-2406729. R.R. acknowledges financial support from the Australian Research Council through DECRA Fellowship DE240100816. This research used resources of the National Energy Research Scientific Computing Center (NERSC), a U.S. Department of Energy Office of Science User Facility located at Lawrence Berkeley National Laboratory, operated under Contract No. DE-AC02-05CH11231.
\end{acknowledgments}

\appendix

\section{EGNN Architecture}\label{sec:appendix_a}
This appendix describes the custom Equivariant Graph Neural Network architecture used in DELTA.  

We consider a graph \( \mathcal{G} = (\mathcal{V}, \mathcal{E}) \), where \( \mathcal{V} \) represents the set of nodes and \( \mathcal{E} \) represents the set of edges. GNNs rely on the concept of message passing, whereby a node accumulates information from its neighbors based on node and edge features. Iterations of message passing causes information to propagate between and accumulate on nodes, building up a rich description of the graph's structure. In the following, subscript $i$ indexes the ``receive'' node and subscript $j$ indexes the ``send'' node, and subscript $ij$ is indicative of an edge between the two. For each receive node, we determine the indices of the send nodes by computing the $k$ nearest neighbors, and represent the set of neighbors connected to $i$ as $N_i$. Each node has:
\begin{itemize}
    \item A spatial position \( \mathbf{x}_i \in \mathbb{R}^3 \).
    \item A feature vector \( \mathbf{h}_i \in \mathbb{R}^d \).
\end{itemize}
To initialize this feature vector, we project the raw node properties $\textbf{p}_i$ into the higher dimensional space for each node using a simple linear node embedding $f_0$, 
\begin{equation}
    \textbf{h}_i^{0}=f_0(\textbf{p}_i).
\end{equation}
Each edge has a relative position vector
\begin{equation}
    \mathbf{r}_{ij} = \mathbf{x}_i - \mathbf{x}_j,
\end{equation}
from which we compute a relative distance with a Euclidian norm,
\begin{equation}
    \quad d_{ij} = \| \mathbf{r}_{ij} \|.
\end{equation}
For the message passing GNN operation we compute edge features using the sending and receiving node features, as well as the distance between the nodes,
\begin{equation}
    \mathbf{e}_{ij}^l= f_e(\mathbf{h}_i^l, \mathbf{h}_j^l, d_{ij}),
\end{equation}
where \( f_e \) is an edge-wise fully connected neural network. Here superscript $l$ indexes the message passing layer, up to $L$ layers. We then aggregate the incoming messages as
\begin{equation}
    \mathbf{m}_i^l = \sum_{j \in N_i} \mathbf{e}_{ij}^l.
\end{equation}
For each layer of the message passing, we update the node features as
\begin{equation}
    \mathbf{h}_i^{l+1} = \mathbf{h}_i^l + f_h(\mathbf{h}_i^l, \mathbf{m}_i^l),
\end{equation}
where \( f_h \) is a node-wise fully connected neural network. Adding to $\textbf{h}$ instead of over-writing at each layer helps stabilize training. After several layers of message passing, the node features accumulate relevant spatial information from nearby nodes. From these accumulated features, we compute a weighting vector for each neighbor based on the final edge features $\textbf{e}_{ij}^L$ 
\begin{equation}
    \textbf{w}_{ij} = f_w(\mathbf{e}_{ij}^L),
\end{equation}
where \( f_w \) is an fully connected neural network that determines the weighting coefficients. We apply these weights to a weighted sum over relative positions to produce an equivariant latent vector for each node,
\begin{equation}
    \mathbf{z}_i =\sum_{j \in N_i} \mathbf{r}^{(2)}_{ij} \cdot \textbf{w}_{ij}.
\end{equation}
Here, $\mathbf{r}^{(2)}_{ij}$ are the spin-2 displacements which enforce the spin-2 rotational symmetry in the $xy$-plane expected from galaxy shapes. These are computed as
\begin{align}
    \bar\theta_{ij} &= \arctan2\left(r_{ij,2}, r_{ij,1}\right) \\
    \bar\phi_{ij} &= \arctan2\left(r_{ij,3}, \sqrt{r_{ij,1}^2 + r_{ij,2}^2}\right)
\end{align}
\begin{equation}
    \mathbf{r}^{(2)}_{ij} = d_{ij} \cdot \begin{bmatrix} \cos(2\bar\theta_{ij}) \\ \sin(2\bar\theta_{ij}) \\ \bar\phi_{ij} \end{bmatrix}.
\end{equation}
Since the weighting is rotationally invariant and the displacements are (spin-2) equivariant, the sum result is a 3D latent space of vectors for each node that are equivariant to rotations in the $xy$-plane. These latent vectors capture the features of each galaxy's surrounding galaxy distribution that are relevant to tidal ellipticity prediction.

\end{document}